\documentclass[12pt,preprint]{aastex}
\usepackage{epsfig}
\usepackage{amssymb}
\usepackage{graphicx}

\slugcomment{To appear in ApJ Letters}

\shorttitle{ALMA observations of Elias 2-24}
\shortauthors{Cieza, L.}

\begin{document}

\title{ALMA observations of Elias 2-24: a protoplanetary disk with multiple gaps in the Ophiuchus Molecular Cloud }

\author{
Lucas A. Cieza\altaffilmark{1,2}, 
Simon Casassus\altaffilmark{3,2}, 
Sebastian P\'erez\altaffilmark{3,2}, 
Antonio Hales\altaffilmark{4,5}, 
Miguel C\'arcamo\altaffilmark{6}, 
Megan Ansdell\altaffilmark{7}, 
Henning Avenhaus\altaffilmark{8},
Amelia Bayo\altaffilmark{9,10}, 
Gesa H. -M. Bertrang\altaffilmark{3,2},
Hector C\'anovas\altaffilmark{11},
Valentin Christiaens\altaffilmark{3,2},
William Dent\altaffilmark{4}, 
Gabriel Ferrero\altaffilmark{12},
Roberto Gamen\altaffilmark{12},
Johan Olofsson\altaffilmark{9,10},
Santiago Orcajo\altaffilmark{12},
Axel Osses\altaffilmark{13},
Karla Pe\~na-Ramirez\altaffilmark{14},
David Principe\altaffilmark{15},
Dary Ru\'iz-Rodr\'iguez\altaffilmark{16},
Matthias R. Schreiber\altaffilmark{9,10},
Gerrit van der Plas\altaffilmark{17},
Jonathan P. Williams\altaffilmark{18},
Alice Zurlo\altaffilmark{1,2} 
}

\newpage
\altaffiltext{1}{Facultad de Ingenier\'ia y Ciencias, N\'ucleo de Astronom\'ia, Universidad Diego Portales, Av. Ejercito 441. Santiago, Chile}
\altaffiltext{2}{Millennium Nucleus ``Protoplanetary Disks in ALMA Early Science", Av. Ejercito 441. Santiago, Chile}
\altaffiltext{3}{Departamento de Astronom\'ia, Universidad de Chile, Casilla 36-D Santiago, Chile}
\altaffiltext{4}{Joint ALMA Observatory, Alonso de Cordova 3107, Vitacura 763-0355, Santiago, Chile}
\altaffiltext{5}{National Radio Astronomy Observatory, 520 Edgemont Road, Charlottesville, Virginia, 22903-2475, USA}
\altaffiltext{6}{Departamento de Ingenier\'ia Inform\'atica, Universidad de Santiago de Chile, Av. Ecuador 3659, Santiago, Chile}
\altaffiltext{7}{Department of Astronomy, University of California at Berkeley, Berkeley, CA 94720-3411}
\altaffiltext{8}{ETH   Zurich,  Institute for Particle Physics and Astrophysics,   Wolfgang-Pauli-Strasse 27, CH-8093, Zurich, Switzerland}
\altaffiltext{9}{Facultad de Ciencias, Instituto de F\'isica y Astronom\'ia, Universidad de Valpara\'iso, Av. Gran Breta–a 1111, 5030 Casilla, Valpara\'so, Chile}
\altaffiltext{10}{N\'ucleo Milenio Formaci\'on Planetaria - NPF, Universidad de Valpara\'iso, Av. Gran Breta\~na 1111, Valpara\'iso, Chile}
\altaffiltext{11}{European Space Astronomy Centre (ESA), Camino Bajo del Castillo s/n, 28692, Villanueva de la Ca\~nada, Madrid, Spain}
\altaffiltext{12}{Instituto de Astrof\'isica de La Plata y Facultad de Ciencias Astron\'omicas y Geof\'isicas, Universidad Nacional de La Plata, Paseo del Bosque s/n, La Plata, Argentina}
\altaffiltext{13}{El Departamento de Ingenier\'ia Matem\'atica, Facultad de Ciencias F\'isicas y Matem\'aticas,  Universidad de Chile, Beauchef 851, Santiago, Chile}
\altaffiltext{14}{Unidad de Astronom\'ia de la Universidad de Antofagasta, Av. U. de Antofagasta. 02800 Antofagasta, Chile}
\altaffiltext{15}{Massachusetts Institute of Technology, Kavli Institute for Astrophysics, Cambridge, MA, USA}
\altaffiltext{16}{Chester F. Carlson Center for Imaging Science, Rochester Institute of Technology, Rochester, NY 14623-5603, USA}
\altaffiltext{17}{Univ. Grenoble Alpes, CNRS, IPAG, F-38000 Grenoble, France}
\altaffiltext{18}{Institute for Astronomy, University of Hawaii at Manoa, Honolulu, HI 96822, USA}

\begin{abstract}
\noindent We present  ALMA 1.3 mm continuum observations at 0.2$''$ (25 au)  resolution of Elias~2-24, one 
of the largest and brightest protoplanetary disks in the Ophiuchus Molecular Cloud, and report the presence of three partially resolved concentric gaps located at $\sim$20, 52, and 87 au from the star. 
We perform radiative transfer modeling of the disk to constrain its surface density and temperature radial profile and place the disk structure in the context of mechanisms capable of forming narrow gaps such as condensation fronts and dynamical clearing by actively forming planets.
In particular, we estimate the disk temperature at the locations of the gaps to be  23,  15, and 12~K  (at  20, 52, and 87~au respectively),  very close to the expected snow-lines of CO %
(23-28 K) and N$_{2}$ (12-15 K). 
Similarly, by assuming that the widths of  the gaps correspond to 4--8 $\times$ the Hill radii of forming planets (as suggested by numerical simulations),  
we estimate planet masses in the range of 0.2--1.5 $M_{\mathrm{Jup}}$, 1.0--8.0 $M_{\mathrm{Jup}}$, and 0.02--0.15 $M_{\mathrm{Jup}}$ for the inner, middle, and outer gap, respectively. 
Given the surface density profile of the disk, the amount of ``missing mass" at the location of each one of  these gaps (between 4 and 20 $M_{\mathrm{Jup}}$) is more than sufficient to account for the formation of such planets. 
\end{abstract}


\section{Introduction}

Gas-rich circumstellar disks are the sites of planet formation. Since most of them are found still embedded in molecular clouds, only a handful of protoplanetary disks are located within 100 pc of Earth.
Given their characteristic temperature and size  (T = 20 K and $r$~$\lesssim$ 100 au;  Williams $\&$ Cieza, 2011), resolving their thermal emission is best achieved by  observations at (sub)millimeter wavelengths with sub-arcsecond angular resolution.  
Imaging protoplanetary disks in great detail has been one of the main scientific drivers for building 
the Atacama Large Millimeter/Submillimeter Array (ALMA), which in recent years has revolutionized our view of protoplanetary disks thanks to its unprecedented resolution.  Before ALMA,  few structures were seen within these disks, mostly in the form of large central cavities, tens of au in radius (e.g.,  Brown et al. 2009;  Andrews et al. 2010; Cieza et al. 2012). When observed at higher resolution and sensitivity,  protoplanetary disks show a variety of substructures such as narrow gaps (ALMA Partnership 2015; Andrews et al. 2016; Isella et al.   2016 ), bright rings (Canovas et al. 2016; van der Plas et al. 2017), dust traps (Casassus et al. 2013; van der Marel et al. 2013; Kraus et al. 2017), spiral arms ( P\'erez et al. 2016), and sharp intensity breaks (Cieza et al. 2016).   The origin of these structures and their role in disk evolution and planet formation is currently one of the major questions in the field. 

Here we present new band-6 (1.3 mm) observations  of Elias 2-24, a well studied K5 star (Prato et al. 2003) hosting one of the largest  and brightest protoplanetary disks  in the nearby Ophiuchus Molecular Cloud (distance $\sim$125 pc; Loinard et al. 2008).  Andrews et al. (2010)  observed this object with a 0.65$''$ $\times$ 0.51$''$ beam at  880 $\mu$m  and detected disk emission up to $\sim$1$"$ from the star with a total flux of 890 mJy.   The disk is close to face-on ($i$ $\sim$25 deg) and
has an estimated  mass of $\sim$0.12 M$_{\odot}$, while the central object is a heavily accreting T Tauri star (\.M$_{\star}$  $\sim$ 2 $\times$ 10$^{-7}$ M$_{\odot}$ yr$^{-1}$; Natta et al. 2006) with a  mass of $\sim$1.0 M$_{\odot}$ and an estimated age of just 0.4 Myr (Andrews et al. 2010; Siess et al. 2000). 
We observed Elias 2-24  as part of the Ophiuchus DIsk Survey Employing ALMA (ODISEA, Cieza et al. in prep.), a program  studying 
147 Ophiuchus  objects at 0.2$''$  (25 au) resolution.
The size, brightness, and orientation of its disk, allow us to search for substructure in Elias ~2-24 using what now can be considered modest resolution. 

\section{ALMA Observations and data analysis}\label{observations}

\subsection{1.3 mm observations}

Elias~2-24 was observed by ALMA in band-6 (230~GHz/1.3 mm) as part of
the Cycle-4 program 2016.1.00545.S on July 13$^{th}$ and 14$^{th}$
2017, in three different execution blocks. 
The precipitable water vapor (PWV) ranged from 1.1 to 1.9~mm in the
three different observing sessions.  During the observations, 42-45
of the 12-m ALMA antennas were used with baselines ranging from 16.7
to 2647.3~m.
The ALMA correlator was configured with two spectral windows with
1.875~GHz bandwidths for continuum observations centered at 232.6 and
218.0~GHz. Also, three spectral windows were placed to cover the
$^{12}$CO (2--1), and $^{13}$CO (2--1), C$^{18}$O (2--1)  transitions of
carbon monoxide at 230.5380, 220.3987, and 219.5603~GHz respectively.
The first spectral window has a 0.04 km s$^{-1}$ spectral resolution, while the other two have a 0.08 km s$^{-1}$ resolution.
J1517-2422 and J1733-1304 served as flux calibrators, while the
quasars J1517-2422 and J1625-2527 were observed for bandpass and phase
calibration respectively. The total integration time on Elias 2-24 was 45 s.

\subsection{Data analysis}

All the ODISEA data were calibrated using the Common Astronomy Software Applications package (CASA v4.2.1;  McMullin et al. 2007) by the ALMA observatory, including the 
the standard bandpass, phase, and amplitude calibrations, the  offline Water Vapor Radiometer (WVR) calibration,  and system temperature corrections. 
Online and nominal flagging, such as shadowed antennas and band edges, were applied for calibration. 
The observations from all  3 execution blocks were concatenated and processed together to increase the signal to noise and uv-coverage. 
We also used CASA to image all ODISEA data using the standard {\tt clean} algorithm,  
with uniform weightings. 
%
The cleaning resulted in a synthesized beam of 0.20$''$ $\times$ 0.25$''$ and a continuum rms of 0.26 mJy beam$^{-1}$. 
After visual inspection, Elias~2-24 immediately stood out as one of the brightest and largest disks of the 147 objects in the survey, and showed a clear ring pattern, with gaps at $\sim$0.4$''$ and $\sim$0.7$''$ (see Figure 1a).
Applying one iteration of phase-only self-calibration reduced the noise level to 0.19 mJy beam$^{-1}$ for this object, 
consistent with the expected thermal noise. 
The CO line was detected with a peak emission of $\sim$60 mJy/beam in 0.25 km/s channels and show rotation broadly consistent with a Keplerian disk, but is not further discussed in this paper. 
The $^{13}$CO and C$^{18}$O lines were not detected. 

Provided with a high dynamic range ($>$200) in the continuum, we investigated the  super-resolution of the visibilities with non-parametric image synthesis. We used the {\tt uvmem} package (Casassus et al. 2006, 2015)  to fit a model image to the data in a least  $\chi^2$  sense.
%
%
In Figure~1b we show a deconvolved model image $\{I_i\}$
using a measure of entropy regularization and the following
objective function: $L = \chi^2 - \lambda S$, with $\chi^2 = \sum_k
\omega_k \| V^\circ_k - V^m_k \|^2$ and $S = \sum_{i} I_i
\log{\biggl(\frac{I_i + G *(\eta + 1)}{G}\biggr)}$. Here $V^\circ$ and
$V^m$ are the observed and model visibility data, $\lambda= 0.1$,
$G=30$, and $\eta = 1.0$.
We also produce a  pure $\chi^2$ model (Figure 1d) with positivity regularization ($I_i > 0$),   which allows for finer angular resolution at the expense of lowering the signal to noise ratio.
The uncertainties in the model are calculated with Monte Carlo simulations, specifically 100 different injections of gaussian random noise in 
the visibilities  (see C\'arcamo et al. 2017 for technical details regarding {\tt uvmen}).
 %
%
%
%
These {\tt uvmem} images have a pixel size of 0.02$''$ and a variable resolution (in the 0.1-0.2$''$ range, depending on the local signal to noise) that is a factor of $\sim$2-3  higher than in the  {\tt clean} image with uniform weights. 
The {\tt clean}  and  {\tt uvmem} images are qualitatively very similar. The main difference is the depths of the two gaps, which are deeper at higher resolution, suggesting that these are
partially resolved features.  
The integrated continuum flux is 345 mJy $\pm$ 35 mJy, where the error reflects the absolute flux calibration.  An elliptical Gaussian fit to the model image indicates a position angle of 43.4  deg (East of North) and an inclination of 23.6 deg. 
Using this inclination and position angle, we create a higher signal-to-noise image averaging  all pixels along concentric ellipses (Figure~1f). 
%
%
%
%

Figure~2  shows a deprojected radial profile of this average image and a cut along the semi-major axis of the disk in the pure $\chi^2$ {\tt uvmem} image. This radial cut maximizes the resolution of the data and allows us to identify a third gap at 0.18$''$ thanks to the increased resolution of {\tt uvmen} toward the center of the image. 
This third gap survives the different realization of noise injected to the visibilities and it is seen as a change in slope in the deprojected radial average. We thus consider it to be real.
There is a small hint of a fourth gap at 0.9$''$, but it has a  low-significance and it is not further discussed in the paper. 

\section{Radiative transfer modeling}\label{modeling}

In order to derive the mass, temperature and surface density profiles of the Elias 2-24 disk, we perform radiative transfer modeling using the code RADMC-3D (Dullemond et al. 2012).  While the disk has barely resolved substructures in the form of gaps, the large scale structure can be approximated as a continuous disk. 
We adopt the same parameterization, stellar parameters,  dust properties, and gas to dust mass ratio (i.e., 100) used by Andrews et al. (2010). In particular, the surface density profile  of the disk, $\Sigma$(r) is given by:

\begin{equation}
\Sigma (r) = \Sigma_c  \left( \frac{r}{R_c} \right)^{-\gamma} exp \Bigg[ -   \left( \frac{r}{R_c} \right)^{2-\gamma}  \Bigg]
\end{equation}

\noindent where $R_c$ is the disk's characteristic radius.  Similarly, the vertical scale height of the disk as a function of radius $h(r)$ is described as:

\begin{equation}
h (r) = h_{100} \left( \frac{r}{R_{100}} \right)^\psi
\end{equation}

\noindent where h$_{100}$ is the scale height at 100 au and $\psi$ defines how this scale height increases with radius, as expected for a flared disk.
In this context, the structure of the disk can be fully described by 5 free parameters:  
$R_{C}$, $\Sigma_c$, $\gamma$, $h_{100}$, and $\psi$.   To facilitate comparisons to other objects,
$\Sigma_c$ can be replaced by $M_{\rm disk}$ (gas $+$ dust disk mass) by integrating Equation~1. 


The parameter space $\{M_{\rm disk}, \gamma, R_{\rm c}, h_{\rm 100}, \psi\}$ was explored using a Bayesian approach, which is described in more detail in P\'erez et al. (in prep), and is based on the Goodman \& Weare's Affine Invariant  
Markov chain Monte Carlo (MCMC) ensemble sampler (Foreman-Mackay et al. 2013) and the publicly available {\sc  python} module {\em emcee}.
The chains were initialized in a uniform distribution around the parameters reported in Andrews et al. (2010) and we used 100 walkers for 500 steps, which allowed the likelihood function to reach steady maximum values.  
The posterior distributions and best fit parameters obtained from the MCMC exploration are shown in Figure~3.
%
%
Our model parameters agree well (within 1--2-$\sigma$) with those derived by Andrews et al.  (2010) using lower resolution data.
The only exception is the  $\psi$ parameter, which is best constrained by the spectral energy distribution (not included in our modeling). 

The comparison between the model and the observations is shown in Figure~1a-c-e, where the deep gap at $\sim$0.4$"$ is clearly visible in the residual as well as hints of the gaps at $\sim$0.18$''$ and $\sim$0.7$''$.
From our radiative transfer model, we can calculate the  temperature of the disk midplane at the location of the gaps:  $\sim$23~K at 20 au, $\sim$15~K at 52 au, and $\sim$12~K at  87 au (see Figure~2).
We can also estimate the mass of the ``missing material" at each gap by integrating the surface density profile of the best-fit model over the width of the gap. Since the gaps are only partially resolved, there is a degeneracy between the width and the depth of the gaps. 
%
%
%
We estimate the widths of the gaps by assuming that they are fully evacuated and measuring their equivalent widths in the brightness profile along the major-axis  (blue line  in Figure~2). 
Gaussian fits  to the gaps using the {\tt splot} routine within the  IRAF package noao.onedspec  indicate that the gaps are located at  20$\pm$3 au, 52$\pm$2 au, 87$\pm$3 au, 
and have widths of 6$\pm$2 au,  28$\pm$3 au, and 11$\pm$4 au.  
%
%
Integrating the area under the curve between the edges of each gap, and/or using  the deprojected radial average (gray line in Figure~2,  in the case of the the two outer gaps), 
produce consistent equivalent widths.  
%
From these gap widths, we derive ``missing masses" of  $\sim$4 $M_{\mathrm{Jup}}$,  20 $M_{\mathrm{Jup}}$, and 10 $M_{\mathrm{Jup}}$ for the gaps at 20, 52 and 87 au, respectively.

\section{Discussion}\label{discussion}

ALMA observations at high resolution and sensitivity have revealed concentric gaps in several sources, including HL Tau (ALMA Partnership et al. 2015), TW Hydra (Andrews et al. 2016), and HD 163296 (Isella et al. 2016).  
%
%
However, the origin of these gaps and rings still remains to be established. Several potential explanations have been offered so far, including dynamical clearing by forming planets (Yen et al. 2017; Dong et al. 2017), snow-lines (Zhang et al. 2015; Okuzumi et al. 2016),   magneto-hydrodynamic effects (Ruge et al. 2016, Flock et al. 2017), and  viscous ring-instability (Dullemond $\&$ Penzlin, 2017).   

Distinguishing between the different potential explanations for concentric gaps  in individual objects seems difficult, and might require the measurements of magnetic fields, dust kinematics, more detailed modeling, and/or the direct detection of planets within the gaps. 
Deep high-contrast searches for planets  in the TW Hydra system yield no detections (Ruane et al. 2017), but place limits of 1.2 to 2.5 $M_{\mathrm{Jup}}$ for the mass of the putative planets at the location of the main gaps. 
However, these limits are not enough to rule out the planetary origin of the gaps as less massive planets might be responsible for them (e.g., Dong et al. 2017).
The demographics of the gaps might also shed some light on their origin.  For instance, if gaps are produced by snow-lines, they must be ubiquitous
and their location should  correlate with the temperature and luminosity of the central source. 
On the other hand, if planets are responsible for these gaps, their location should not depend on disk temperature, but  their widths  and depths should correlate with the mass of the disk.
In the case of our Ophiuchus survey (Cieza et al. in preparation), among 147 sources, only the disk around  Elias 2-24 has the right combination of  size, brightness, and orientation that allows the identification of three gaps.  
%
Deeper, higher-resolution images are needed to investigate the overall incidence and properties of concentric gaps in the Ophiuchus Molecular Cloud and other star-forming regions. 

Meanwhile, we can place the results presented in Section~3 in the context of opacity gaps created by condensation fronts and the dynamical clearing by forming planets.  
Zhang et al. (2015) suggest that the most prominent gaps  seen in HL Tau  (at  13, 32, and 63 au) are due to the snow-lines of water ($\sim$144 K) pure ammonia or ammonia hydrates ($\sim$84 K) and  clathrate hydrates ($\sim$57 K). In the case of Elias 2-24, all these temperatures correspond to stellocentric distances $<$ 20 au, which are not resolved by our observations.  The disk temperatures we derive for the location of the gaps (23, 15, and 12 K) are instead very close to the condensation fronts of other species: CO, and N$_2$, which are expected to occur at  temperatures of 23-28 K and 12-15 K (Mumma $\&$ Charnley 2011; Mart\'in-Dom\'enech et al. 2014).  Interestingly,  Andrews et al.  (2016) reports that the gaps seen at 22 and 37 au in TW Hydra are also located close to the expected snow-lines of CO and N$_{2}$ according to their thermal model.  However, the gaps in TW Hydra are narrower than those of Elias 2-24, both in terms of absolute size (1-6 au vs 8 - 20 au) and as a fraction of disk radius ($<$ 2--8 $\%$ vs 6--20 $\%$). 

To date, the closest analog to Elias 2-24 in terms of structure seems to be HD~163296 (Isella et al. 2016). The 1.3 mm images of both objects are remarkably similar, but HD~163296 is actually a scaled-up version (by a factor of two) of Elias 2-24 with respect to the stellar mass, the disk size, the location of the gaps and their widths.  HD~163296 has a dust disk 250 au in radius with three gaps at 60, 100, and 160 au. These gaps have estimated widths of 25, 22, and 45 au respectively.  
Isella et al. (2016) argues that the CO snow-line in HD 163296 could be anywhere between 50 and 180 au and therefore do not associate frost lines with any particular gap. Instead, they estimate the masses of the possible planets that could explain the observed gaps.  They note that numerical simulations show that planets can open gaps that are 4 to 8 times the Hill radius (Wolf et al. 2007; Rosotti et al. 2016; ),  $r_{Hill}$, which is given by:

\begin{equation}
r_{Hill} \sim  a\times \left( \frac{m_p}{3M_{\star}} \right)^{1/3}
\end{equation}

\noindent where $a$ is the semi-major axis of the planet's orbit, $m_p$ is the mass of the planet and M$_{\star}$ is the mass of the central star, 2.3 M${_\odot}$ in the case 
of  HD~163296.
Using this argument,  they estimate that planets with masses between 0.05 and 0.5 $M_{\mathrm{Jup}}$ are needed to explain the gaps in HD 163296. 
Applying the same approach to Elias 2-24, we derive planets masses in the range of 0.2--1.5 $M_{\mathrm{Jup}}$, 1.0--8.0 $M_{\mathrm{Jup}}$, and 0.02--0.15 $M_{\mathrm{Jup}}$ for the gaps at 20, 52 and 87 au, respectively. These values assume the gaps widths adopted in Section~3 (6, 28, and 11 au, respectively) and a stellar mass of 1.0 M$_{\odot}$ (Andrews et al. 2016).
The planet masses so derived are very modest compared to ``missing masses", also reported in Section~3, based on our continuous disk model (4--20 $M_{\mathrm{Jup}}$). We thus conclude that there was more than enough mass at the location of the gaps (as an initial condition) to account for the formation of the putative planets. In this scenario,  most of the ``missing mass" would likely be pushed away by the tidal forces of the planet toward the edges of the gap, while only a small fraction of this mass would be accreted by each planet (Szul\'agyi et al. 2014; Dipierro $\&$ Laibe,  2017).
We note that the two explanations discussed above for the structure of Elias 2-24 (condensation fronts and dynamical clearing by forming planets) might be closely related, as one can imagine combined scenarios in which the  conditions at the condensation fronts are responsible for the first steps of planet formation (the growth of pebbles and planetesimals) that eventually lead to large planets capable of dynamically clearing the gaps. However,  opacity gaps produced by snow-lines would only require the formation of cm-sized  pebbles, while  dynamical clearing would imply the formation of giant planets at $\sim$20-85 au by the age of the system, $<$ 1.0 Myr in the case of Elias 2-24 (Andrews et al.  2010). 
%
%

\section{Summary and Conclusions}

We have obtained ALMA 1.3 mm observations of Elias~2-24  as part of the ODISEA program which is a survey of 147 objects in Ophiuchus at  $\sim$25 au resolution. The disk shows a structure of  three concentric gaps with  estimated widths ranging from $\sim$6 to 28 au.  Radiative transfer modeling of the source indicates that the disk temperature at the location of the gaps is close to the expected snow-lines of CO and N$_{2}$, consistent with claims that frost lines can result in gaps of dust opacity.  The surface density profile of the disk is also consistent with formation of planets with masses similar to those of the giant planets  in the Solar System (from Neptune-mass  to a few times the mass of Jupiter).  However, other potential explanations still exist, including  magneto-hydrodynamic effects  and  disk instabilities.   Elias~2-24 is one of the brightest and largest disks in the Ophiuchus Molecular cloud, and it remains to be established whether smaller versions of these gaps  (e.g., with locations and widths scaled-down by disk radii and/or stellar luminosity)  are typical of protoplanetary disks.  Previous observations of TW Hydra at much higher resolution ($\sim$1 au) suggest that this might be the case. 

\acknowledgments
This paper uses  the following ALMA data: ADS/JAO.ALMA \#2016.1.00545.S.  ALMA is a partnership of ESO, NSF (USA) and NINS (Japan), together with NRC (Canada), NSC and ASIAA (Taiwan), and KASI (Republic of Korea), in cooperation with the Republic of Chile. 
The NRAO  is a facility of the NSF operated under cooperative agreement by Associated Universities, Inc.
L.A.C., S.C., G.H.M.B,  A.O, and A.Z  were supported by CONICYT-FONDECYT grant numbers 1171246,  1171624, 3170204, 1151512, and 3170657. 
L.A.C., S.C., S.P., G.H.M.B.,  and A.Z. acknowledge support from the Millennium Science Initiative (Chile) through grant RC130007. 
A.B., J.O., and M.R.S. acknowledge financial support from ICM N\'ucleo Milenio de Formaci\'on Planetaria, NPF.
%
%
This paper use the Brelka cluster, financed by Fondequip project EQM140101.

\newpage

\begin{figure}
\includegraphics[width=16cm, trim = 0mm 0mm 0mm 0mm, clip]{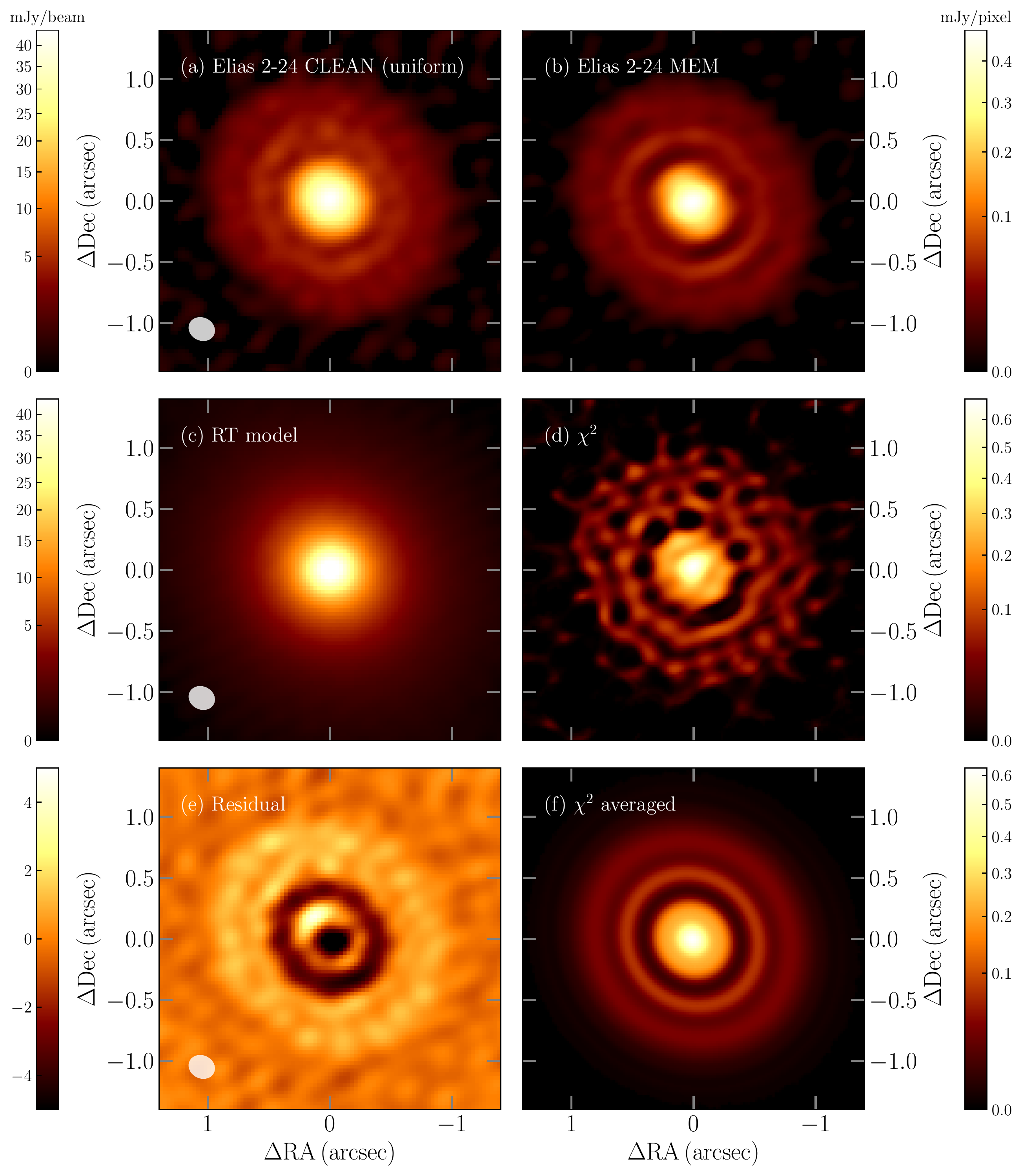}
\caption{ \small 
 The 1.3 mm  {\tt clean} image  of Elias 2-24  with uniform weights (a) and the deconvolved model image using the {\tt uvmem} package with the parameters 
 described in the text (b).
  The best-fitting model (c) and the pure $\chi^2$  {\tt uvmem} image (d). 
 The radiative transfer  residual (e), where the outer gaps are seen as negative features as they are not included in the modeling. The asymmetric structure close to the origin is likely to be due to the inner-most gap, which is only marginally resolved.
 The elliptical average of the pure $\chi^2$    {\tt uvmem} image (f).  
%
}
\label{fig:MEM}
\end{figure}

%
%

\begin{figure}
\includegraphics[width=18cm, trim = 15mm 30mm 0mm 0mm, clip]{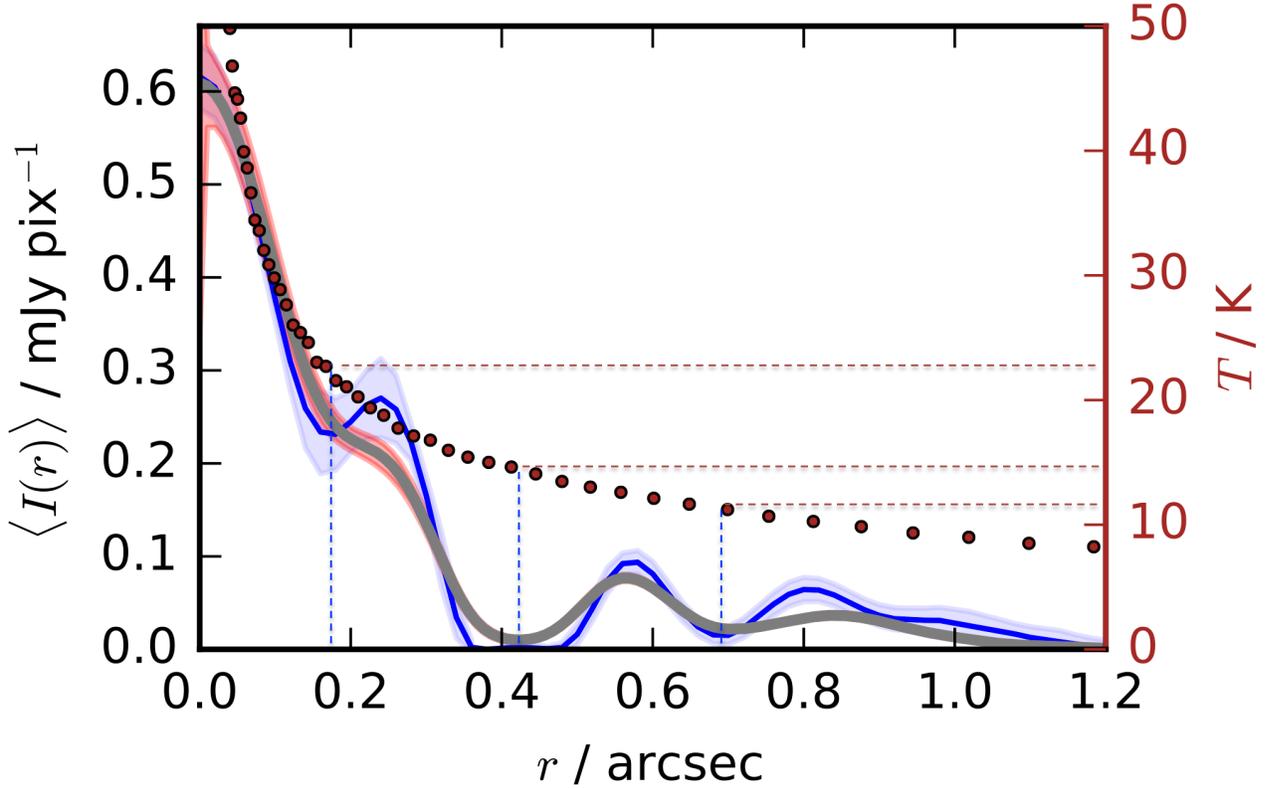}
\caption{ \small 
The radial brightness profile of Elias 2-24 at 1.3 mm. 
The gray line and red shade correspond to the \emph{deprojected} radial average of the image (Figure~1f)
 and the errors in this average. 
The blue line corresponds to a cut along the semi-major axis of the pure $\chi^2$ {\tt uvmem}  image (Figure~1d) and maximizes the resolution of the data.
In this case, the error in the profile (blue shade) is derived from 100 Monte Carlo realizations  of noise injected to the visibilities. 
The vertical lines mark the location of the gaps, which correspond to distances of $\sim$20, 52 and 87 au.   
The dotted line shows the midplane temperature profile of our continuous disk model with the scale on the right, 
while the horizontal lines indicate the approximate temperatures at the location of the gaps. 
}
\label{fig:profiles}
\end{figure}

\begin{figure}
\includegraphics[width=17cm, trim = 0mm 120mm 0mm 0mm, clip]{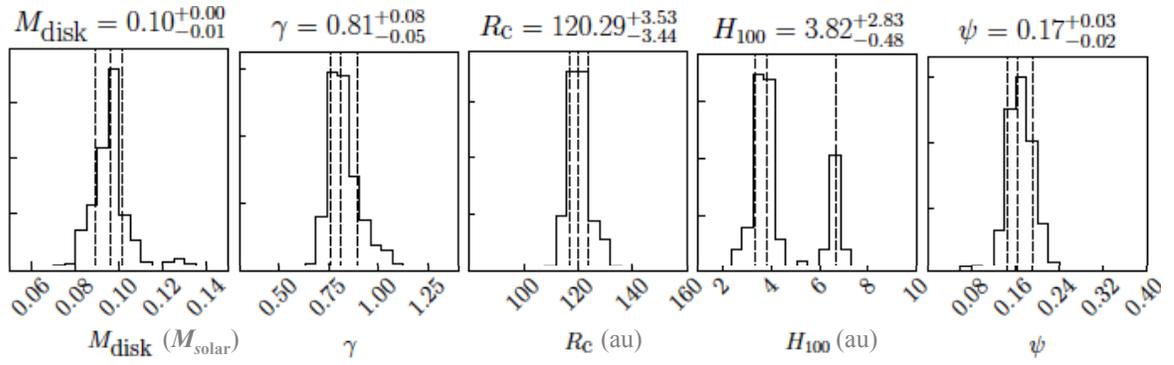}
\caption{ \small 
Posterior distributions of each of the 5 parameters used to model Elias~2-24 as a continuous disk.
The vertical dashed lines represent the 16th, 50th and 84th percentiles. 
}
\label{fig:MEM}
\end{figure}


\end{document}